\documentclass[twocolumn]{aastex631}
\hypersetup{linkcolor=blue,citecolor=blue,filecolor=blue,urlcolor=blue}
\shorttitle{Revisiting oldest stars as cosmological probes}
\shortauthors{Cimatti A. \& Moresco M.}

\begin{document}
\title{Revisiting oldest stars as cosmological probes: new constraints on the Hubble constant}
\author[0000-0002-4409-5633]{Andrea Cimatti}
\affiliation{Dipartimento di Fisica e Astronomia ``Augusto Righi'', Universit\`a di Bologna, Viale Berti Pichat 6/2, I-40127, Bologna, Italy}
\affiliation{INAF - Osservatorio di Astrofisica e Scienza dello Spazio di Bologna, via Gobetti 93/3, 40129 Bologna, Italy}
\author[0000-0002-7616-7136]{Michele Moresco}
\affiliation{Dipartimento di Fisica e Astronomia ``Augusto Righi'', Universit\`a di Bologna, Via Gobetti 93/2, I-40129, Bologna, Italy}
\affiliation{INAF - Osservatorio di Astrofisica e Scienza dello Spazio di Bologna, via Gobetti 93/3, 40129 Bologna, Italy}
\begin{abstract}
Despite the tremendous advance of observational cosmology, the value of the Hubble constant ($H_0$) is still controversial (the so called ``Hubble tension'') because of the inconsistency between local/late-time measurements and those derived from the cosmic microwave background. As the age of the Universe is very sensitive to $H_0$, we explored whether the present-day oldest stars could place independent constraints on the Hubble constant. To this purpose, we selected from the literature the oldest objects (globular clusters, stars, white dwarfs, ultra-faint and dwarf spheroidal galaxies) with accurate age estimates. Adopting a conservative prior on their formation redshifts ($11 \leq z_{\rm f} \leq 30$) and assuming $\Omega_{\rm M} = 0.3 \pm 0.02$, we developed a method based on Bayesian statistics to estimate the Hubble constant. We selected the oldest objects ($>13.3$ Gyr) and estimated $H_0$ both for each of them individually and for the average ages of homogeneous subsamples.  Statistical and systematic uncertainties were properly taken into account. The constraints based on individual ages indicate that $H_0<70.6$ km/s/Mpc when selecting the most accurate estimates. If the ages are averaged and analyzed independently for each subsample, the most stringent constraints imply $H_0<73.0$ with a probability of 90.3\% and errors around 2.5 km/s/Mpc. We also constructed an ``accuracy matrix'' to assess how the constraints on $H_0$ become more stringent with further improvements in the accuracy of stellar ages and $\Omega_{\rm M}$. The results show the high potential of the oldest stars as independent and competitive cosmological probes not only limited to the Hubble constant.
\end{abstract}
\section{Introduction} 
\label{sec:intro}
Our understanding of the Universe improved dramatically during the last century. However, in spite of the high precision achieved in observational cosmology, several fundamental questions remain open. For instance, the nature and origin of dark matter and dark energy are still unknown despite their major contribution to the total cosmic budget of matter and energy \citep{planck2020}. Another key question regards the present-day expansion rate (the Hubble constant, $H_0$), for which independent methods give inconsistent results, e.g. $67.4 \pm 0.5$ km/s/Mpc \citep[][hereafter Planck2020]{planck2020} and  $73.04 \pm 1.04$ km/s/Mpc \citep[][hereafter SH0ES]{riess2022}, leading to the so called ``Hubble tension'' \citep{verde2019,abdalla2022,kamionkowski2022}. Today, it is unknown whether the $H_0$ discrepancy is a signal of ``new physics'' or the result of unaccounted systematic effects. Thus, before adventuring into the uncharted territory of new physics, it is essential to combine as many as possible cosmological probes in order to mitigate the unavoidable systematic uncertainties inherent to each of them \citep[for a review, see][]{moresco2022}.
In this regard, stellar ages play a key role simply because the current age of the Universe today cannot be younger than the age of the present-day oldest stars. Historically, the ages of the oldest globular clusters appeared inconsistent with the mostly younger ages of the Universe allowed by the cosmological models in the 1980s and early 1990s \citep[][and references therein]{jimenez1996,krauss2003}. This age crisis was rapidly solved with the discovery of the accelerated expansion which implied an older Universe. More recently, stellar ages have been reconsidered as promising probes {\em independent of the cosmological models} \citep{bond2013,jimenez2019,valcin2020,valcin2021,boylan2021,vagnozzi2022,weisz2023b}.
As a matter of fact, age dating is based either on stellar physics and evolution (isochrone fitting) or on the abundance of radioactive elements (nucleochronometry) \citep{soderblom2010}. The downside is that stellar ages are still affected by substantial systematic uncertainties \citep[e.g.,][]{chaboyer1995,soderblom2010,valcin2021,joyce2023}. In particular, isochrone fitting relies on the assumption of a given theoretical stellar model and requires accurate estimates of metal abundance, absolute distance, and dust reddening along the line of sight. Thus, although the age precision can be very high for a given set of assumptions (i.e. statistical errors can be very small), high accuracy is usually prevented by systematic errors. In nucleochronometry, an additional difficulty is the accurate derivation of the abundances of elements (e.g. U, Th) characterized by very weak, and often blended, absorption lines \citep[e.g.][]{christlieb2016}.

The main aim of this paper is to revive and investigate the potential of the oldest stars as independent clocks to place new constraints on the Hubble constant. 
\section{Method}
\label{sec:method}
In a generic cosmological model, the Hubble constant $H_0$ can be derived as:
\begin{equation}
H_0 = \frac{A}{t}\int_{0}^{z_f}\frac{1}{(1+z')E(z')}dz'
\label{eq:H0}
\end{equation}
where $E(z)=H(z)/H_0$, $t$ is the age of an object formed at redshift $z_f$ and $A=977.8$ allows to convert from Gyr (units of $t$) to km/s/Mpc (units of $H_0$). For $z_f=\infty$, the age $t$ converges to the age of the Universe $t_U$.
In a flat $\Lambda$CDM universe, Eq.~\ref{eq:H0} reduces to:
\begin{equation}
H_0=\frac{A}{t}\int_{0}^{z_f}\frac{1}{1+z'}\left[\Omega_{\rm M} (1+z')^3+(1-\Omega_{\rm M})\right]^{-1/2}dz'\; .
\label{eq:H0_LCDM}
\end{equation}
Based on Eq.~\ref{eq:H0_LCDM}, it is therefore possible to estimate $H_0$ provided that $\Omega_{\rm M}$, $z_f$, and stellar ages are known. The sensitivity of this method is described in Sect.~\ref{sec:sim}.
\section{The oldest stars in the present-day Universe}
\label{sec:age_meas}
The age of the Universe ($t_{\rm U}$ at $z=0$) is very sensitive to $H_0$. For instance, for $\Omega_{\rm M}=0.3$ and $\Omega_{\Lambda}=0.7$, the age of the Universe is $t_{\rm U} \sim 14.1$ Gyr and $t_{\rm U} \sim 12.9$ Gyr for $H_0=67$ km/s/Mpc and $H_0=73$ km/s/Mpc, respectively. From this example, it is clear that only the {\em oldest} stars play a discriminant role in the context of the Hubble tension. 
With this motivation, we searched the literature for the oldest stars in the Milky Way and in the Local Group with ages estimated based on different methods and with a careful evaluation of systematic errors.

\begin{itemize}
\item {\em Galactic globular clusters (GC)}. For our purpose, we focused on the most recent results of \cite{omalley2017}, \cite{brown2018}, \cite{oliveira2020}, and \cite{valcin2020} with state-of-art age dating and a careful assessment of the statistical and systematic uncertainties. The oldest GCs have ages $\gtrsim 13.5$ Gyr with total errors (i.e. combined statistical and systematic) from $\sim 0.5$ Gyr to $\gtrsim 1$ Gyr. In particular, in the case of \cite{omalley2017} we used the weighted ages based on the most accurate parallaxes available (i.e. the combined ages in their Tab. 5). Finally, we also added the recent estimate of the absolute age of the globular cluster M92 by \cite{ying2023} ($13.80\pm0.75$ Gyr) where a careful analysis of the error budget is also presented. It is important to emphasize that these different works span a variety of statistical methods (Bayesian, Monte Carlo), stellar models (BaSTI, \cite{dotter2008}, \cite{vandenberg2014}), ranges of fitted parameters (age, [Fe/H], [$\alpha$/Fe], distance, reddening), sometimes exploiting {\em Gaia} parallaxes.
\item {\em Galactic individual stars}. Very old individual stars are reported in the literature. For instance, \cite{schlaufman2018} estimated an age of 13.5 Gyr (with a systematic error $\gtrsim 1$ Gyr) for an ultra metal-poor star belonging to a binary system. For HD 140283, an extremely metal-poor star in the solar neighborhood, an age of $14.5 \pm 0.8$ Gyr (including systematic uncertainties) was derived by \cite{bond2013}, although recent results suggest younger ages \citep{plotnikova2022}. Recent works (based on Gaia parallaxes, sometimes with asteroseismology measurements and without adopting priors on the age of the Universe) found evidence of stars with ages $\gtrsim 13.5$ Gyr \citep[e.g.,][]{montalban2021,xiang2022, plotnikova2022}. Regarding the data from \cite{plotnikova2022}, we used the ages estimated with BaSTI isochrones because they provide a larger dataset.
\item {\em White dwarfs (WDs)}. If the distance, magnitude, color, and atmospheric type for a WD are known, its age can be derived based on the well-understood WD cooling curves and initial–final mass relations calibrated using star clusters. \cite{fouesneau2019} exploited the Gaia parallaxes and reported ages as old as $13.9 \pm 0.8$ Gyr. The potential of WDs as chronometers has been recently highlighted by \cite{moss2022}.
\item {\em Nucleochronometry}. The relative abundances of nuclides with half-lifes of several Gyr (e.g. U, Th, Eu) can be exploited as chronometers \citep{christlieb2016,shah2023}. However, its application requires reliable theoretical modeling of the rapid neutron capture (r-process) nucleosynthesis and spectroscopy with very high resolution and signal-to-noise ratio. To date, this method has been applied only to a few stars whose ages turned out to be as old as $\approx 14$ Gyr, but with large errors of $2-4$ Gyr. However, \cite{wu2022} suggested that the uncertainties could be reduced down to $\sim 0.3$ Gyr through the synchronization of different chronometers.
\item {\em Ultra faint galaxies (UFDs) and dwarf spheroidals (dSph)}. UFDs in the Local Group have old stellar populations and may be the fossil remnants of systems formed in the reionization era. \cite{brown2014} found that the oldest stars have ages in the range of $13.7-14.1$ Gyr, with systematic uncertainties of $\sim 1$ Gyr. Moreover, based on the reconstruction of their star formation histories, some dSph systems of the Local Group formed the bulk of their stars at $z>14$, therefore implying ages $>$13.5 Gyr in the standard $\Lambda$CDM cosmology \citep[e.g.,][]{weisz2014, simon2022}. 
\end{itemize}

The oldest ages selected for our work are based on a variety of astrophysical objects, methods and independent studies, and show unambiguously that the most ancient stars in the present-day Universe are significantly older than 13 Gyr, but with uncertainties (dominated by systematic errors) from $\sim 0.5$ to $\gtrsim 1$ Gyr.

Can such old ages place meaningful cosmological constraints? We recall the obviousness that the age of an object at $z=0$ provides only a lower limit to the current age of the Universe as it remains unknown how much time it took for that object to form since the Big Bang:
\begin{equation}
t_{\rm U} = \Delta t_{\rm f} +  t_{\rm age}
\label{eq:tU}
\end{equation}
where $t_{\rm U}$ is the age of the Universe, $\Delta t_{\rm f}$ is the time interval between the Big Bang and the formation of an object observed at $z=0$ with an age $t_{\rm age}$. Thus, if we measure $t_{\rm age}$ for an object at $z=0$, the main unknown remains only $\Delta t_{\rm f}$. 
In our work, we exploited the {\em oldest} stars at $z=0$ to maximize $t_{\rm age}$ and minimize the relevance of $\Delta t_{\rm f}$ with respect to the current age of the Universe. To this purpose, we decided to anchor $\Delta t_{\rm f}$ to the redshifts ($z \sim 11-13$) of the most distant galaxies known based on spectroscopic identification \citep{curtislake2022}, although photometric candidates exist up to $z \approx 18$ \citep{naidu2022}. Our choice is also indirectly supported by the recent discovery of high-redshift quiescent galaxies whose star formation histories require that their first stars formed at $z>11$ \citep[see, e.g.,][]{carnall2023}. The uppermost redshift limit can be set by theoretical models that indicate $20<z<30$ as the range for the formation of the very first stars \citep{galli2013}. 
Thus, for our analysis (Sect.~\ref{sec:method}), we adopted $11<z_f<30$ as a baseline. This corresponds to $\Delta t_{\rm f}\approx 0.1-0.4$ Gyr after the Big Bang ($H_0 = 70$ km/s/Mpc, $\Omega_{\rm M}=0.3$, $\Omega_{\Lambda}=0.7$). 
We remark that this choice is the most conservative possible for the Hubble tension: should the oldest stars have formed at $z<11$, their ages would imply an even older universe and, in turn, a lower value of $H_0$.
\section{Constraining the Hubble constant}
\label{sec:cosmoconstraints}

For our analysis, we developed a code based on a Bayesian framework, with a log-likelihood defined as:
\begin{equation}
\mathcal{L}({\rm age},\mathbf{p})= - 0.5 \sum_i\frac{\left({\rm age}_{i}-age_m(\mathbf{p})\right)^2}{\sigma({\rm age}_i)}
\end{equation}
where $age_i$ and $\sigma({\rm age}_i$) are the age and its error, $age_m$ is the theoretical age from the model in Eq.~\ref{eq:H0_LCDM}, and $\mathbf{p}$ are the parameters of the model. We adopted a flat $\Lambda$CDM cosmological model where the free parameters are ($H_0$, $\Omega_{\rm M}$, and $z_{\rm f}$). We sampled the posterior with a Monte-Carlo Markov Chain approach using the affine-invariant ensemble sampler implemented in the public code \texttt{emcee} \citep{foreman2013}.

While we decided to adopt flat priors on $H_0=[50,100]$ and $z_{\rm f}=[11-30]$, we chose to include a Gaussian prior on $\Omega_{\rm M}$ because, as can be inferred from Eq.~\ref{eq:H0_LCDM}, there is a significant intrinsic degeneracy between the derived value of $H_0$ and $\Omega_{\rm M}$ that can be hardly broken from age data alone. Most importantly, in the framework of the current Hubble tension \citep[see, e.g.,][]{verde2019}, we kept our analysis free from CMB-dependent priors by adopting $\Omega_{\rm M}=0.3\pm 0.02$ obtained from the combination of several low-redshift results \cite{jimenez2019}, and consistently with the latest BOSS+eBOSS clustering analysis ($\Omega_{\rm M}=0.29^{+0.012}_{-0.014}$) of \cite{semenaite2022}. 
In our analysis, we adopted 250 walkers and 1000 iterations each, discarding the first 300 points of the chain to exclude burn-in effects.

\section{Sensitivity to the choice of priors}
\label{sec:sim}

\begin{table}[t!]
\centering
\begin{tabular}{c|c}
\hline
\hline
parameter & values and priors \\
\hline
Age [Gyr] & 12.5, 13, {\bf 13.5}, 14\\
$\sigma(age)$ [Gyr] & 0.1, 0.2, {\bf 0.5}, 1, 1.5, 2\\
$z_{f}$ & {\bf [11-30]}, [15-30], [20-30]\\
$\Omega_{\rm M}$ & $\mathbf{0.3\pm0.02}$, $0.27\pm0.06$ \citep{verde2002},\\
 & $0.341\pm0.045$ \citep{gilmarin2017}\\
\hline
\end{tabular}
\caption{Range of parameters and priors explored to assess the sensitivity of the method. The {\it reference case} discussed in the text is indicated in boldface.}
\label{tab:priors}
\end{table}

We first explored the sensitivity of the method by fitting a theoretical grid of parameters, spanning different values of ages and errors, and assessing the impact of changing the priors of $z_f$ or $\Omega_{\rm M}$ on $H_0$. To this purpose, we defined a \textit{reference case} with age=$13.5\pm0.5$ Gyr, a flat prior $z_f=[11-30]$, and a Gaussian prior $\Omega_{\rm M}=0.3\pm0.02$. 

\begin{figure}[t!]
\centering
\includegraphics[width=0.47\textwidth]{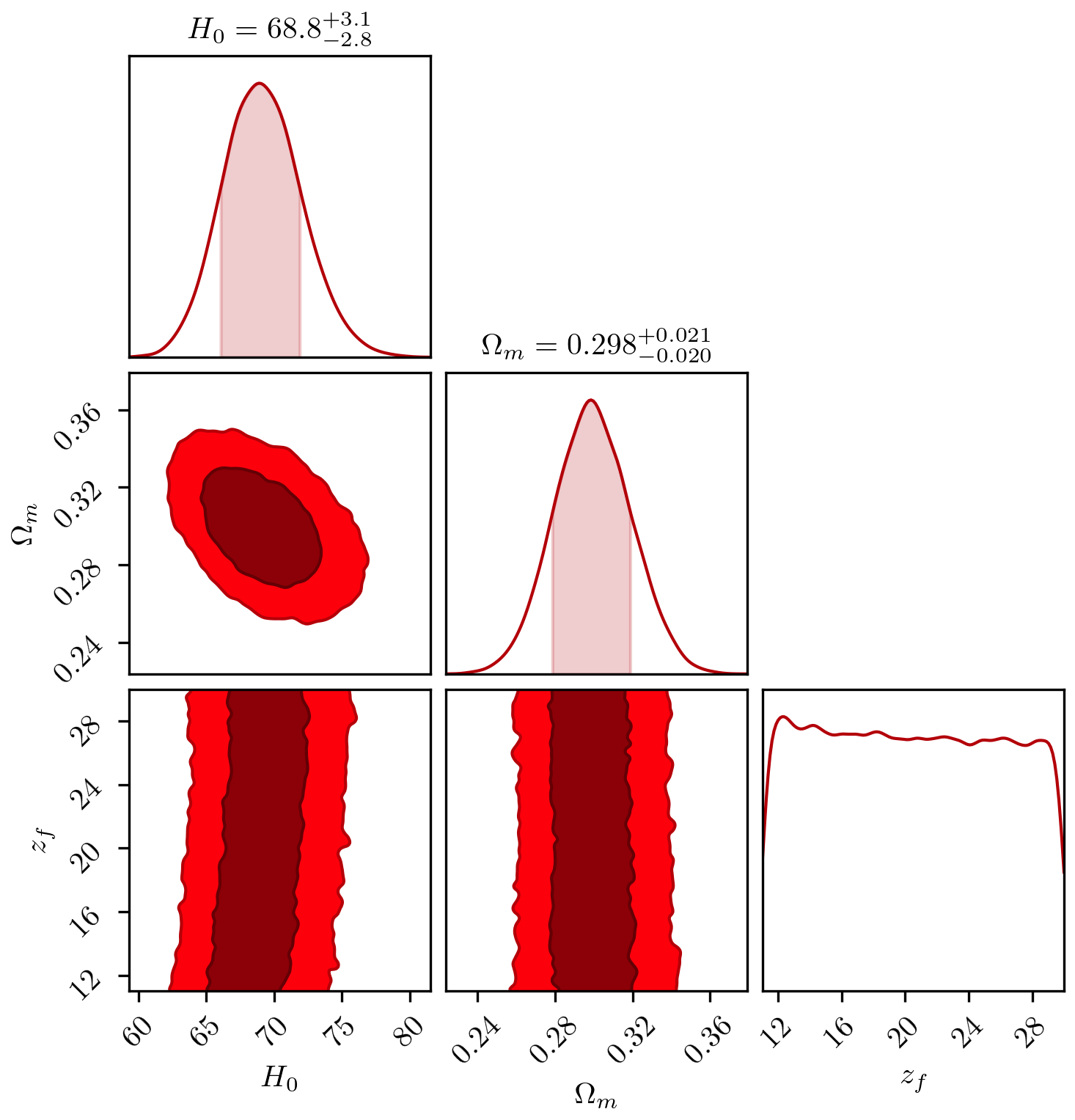}
\includegraphics[width=0.49\textwidth]{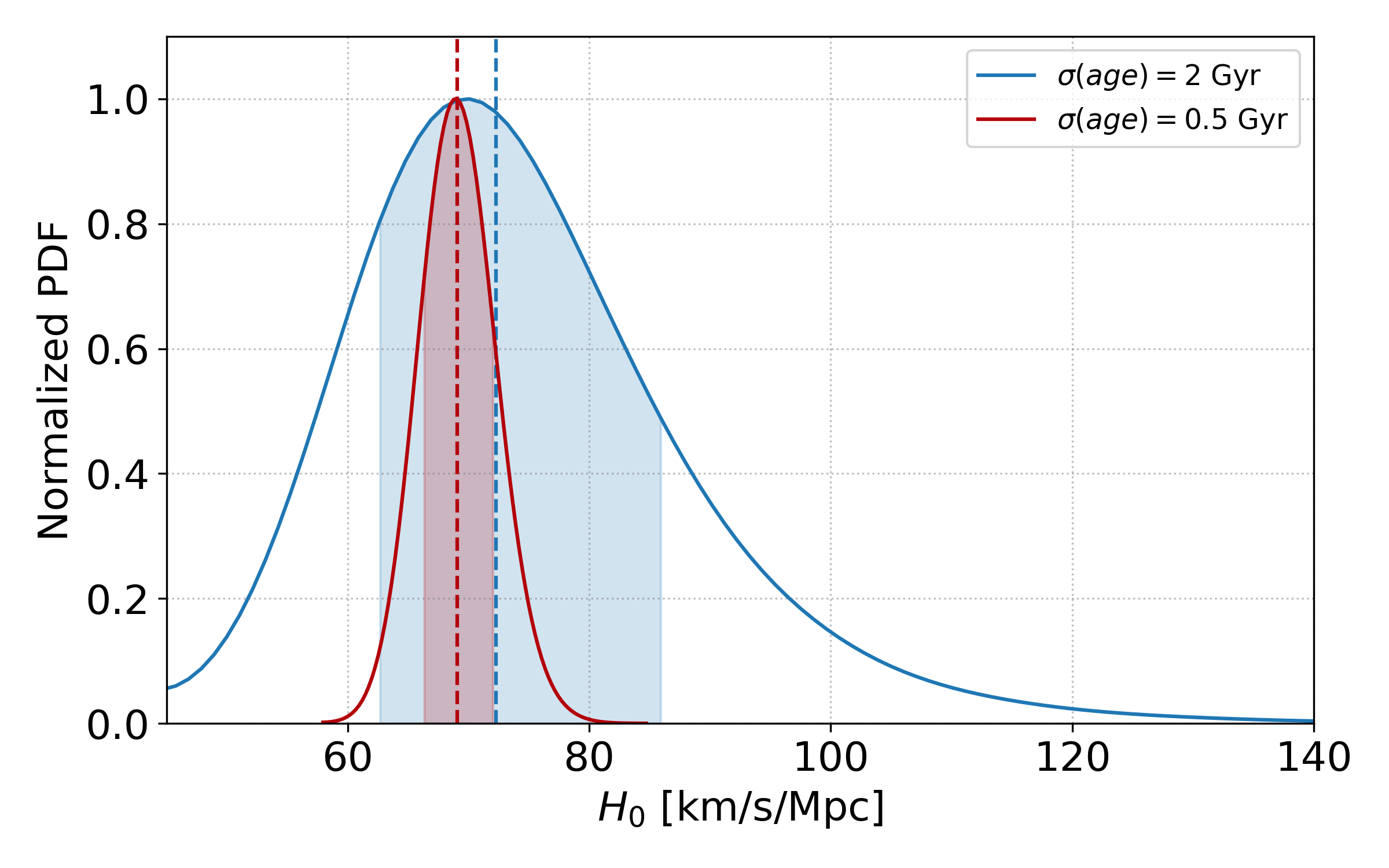}
\caption{\textit{Upper panel.} Constraints on $H_0$ for the \textit{reference case} with an age of $13.5\pm0.5$ Gyr, a flat prior of $z_f=[11-30]$ and a Gaussian prior on $\Omega_{\rm M}=0.3\pm0.02$. The results have a very weak dependence on $z_f$ (showing a flat distribution), as also confirmed from the analysis in Sect.~\ref{sec:sim}. \textit{Lower panel.} The effects of different age errors on the $H_0$ estimate. The two curves show the normalized PDF for the \textit{reference case}, but with two different age errors as indicated in the label; the shaded area represents the 1-$\sigma$ confidence level. For increasingly large errors on the age, the PDF becomes progressively more asymmetric biasing the estimate of $H_0$ towards high values, as indicated by the median of the distributions (vertical dashed lines).}
\label{fig:H0}
\end{figure}

We then perturbed the \textit{reference case} by changing, each time, one of the assumed values or priors within the ranges indicated in Tab.~\ref{tab:priors}. 
The results for the \textit{reference case} are shown in Fig.~\ref{fig:H0}. 

The general findings can be summarized as follows.
\begin{itemize}
\item As expected, the estimated value of $H_0$ decreases linearly with increasing age, varying from 69 to 65 km/s/Mpc for ages from 13.5 to 14.5 Gyr (for the fixed priors in Tab.~\ref{tab:priors}). We note that to obtain a Hubble constant $>$73 km/s/Mpc, the oldest stars in the Universe should be at most 12.75 Gyr old for the assumed priors on $\Omega_{\rm M}$, or, alternatively, the matter density parameter should be $\Omega_{\rm M}<0.25$ for an age of $\sim$13.5 Gyr.
\item The uncertainty on the Hubble constant scales almost linearly with the uncertainty on the age. For $\sigma_{age}\lesssim0.5$ Gyr, the probability distribution function of $H_0$ (PDF; Fig.~\ref{fig:H0} bottom panel, red curve) is Gaussian. However, the PDF becomes asymmetric for $\sigma_{age}\gtrsim1$ Gyr with tails that bias the estimate of $H_0$ towards larger values (Fig.~\ref{fig:H0} bottom panel, blue curve).
\item The $H_0$--$\Omega_{\rm M}$ degeneracy plays an important role. For the three cases of $\Omega_{\rm M}$ shown in Tab.~\ref{tab:priors}, the resulting values of $H_0$ are $71.32^{+5.8}_{-4.76}$, $69.06^{+2.96}_{-2.77}$, and $66.6^{+2.64}_{-2.42}$ km/s/Mpc for $\Omega_{\rm M}=0.27 \pm 0.06$, $0.30\pm0.02$ and $0.34\pm0.045$. The uncertainty on $H_0$ would decrease to $\pm2.5$ km/s/Mpc assuming the Planck2020 value ($\Omega_{\rm M}=0.315\pm0.007$). The systematic uncertainty due to the range of $\Omega_{\rm M}$ priors in Tab.~\ref{tab:priors} is $\sigma_{syst,\;\Omega_{\rm M}}=2.36$ km/s/Mpc.
\item The prior on $z_f$ plays a minor role. For the ranges of $z_f$ in Tab.~\ref{tab:priors}, the systematic uncertainty is only $\sigma_{syst,\;z_f}=0.25$ km/s/Mpc. As a further example, adopting $6<z_f<11$,  age=$13.5\pm0.5$ Gyr and $\Omega_{\rm M} = 0.30\pm0.02$, $H_0$ would decrease to $66.97^{+2.94}_{-2.76}$ compared to $69.09^{+2.96}_{-2.75}$ of the case with $11<z_f<30$.
\end{itemize}
Based on this analysis, the total systematic error associated with the prior assumptions is $\sigma_{syst,\;prior}(H_0)=\sqrt{\sigma_{syst,\;z_f}^2+\sigma_{syst,\;\Omega_{\rm M}}^2}=2.37$ km/s/Mpc. However, this is a conservative additional systematic error as the range of our priors already covers a parameter space well constrained by observations (see Sect.~\ref{sec:age_meas}).

\section{From the oldest ages to \texorpdfstring{$H_0$}{H0}}
\label{sec:H0}
The method described in Sect.~\ref{sec:cosmoconstraints} was then applied to the observed data. In particular, we followed two different approaches.

\begin{figure*}[t!]
\centering
\includegraphics[width=0.95\textwidth]{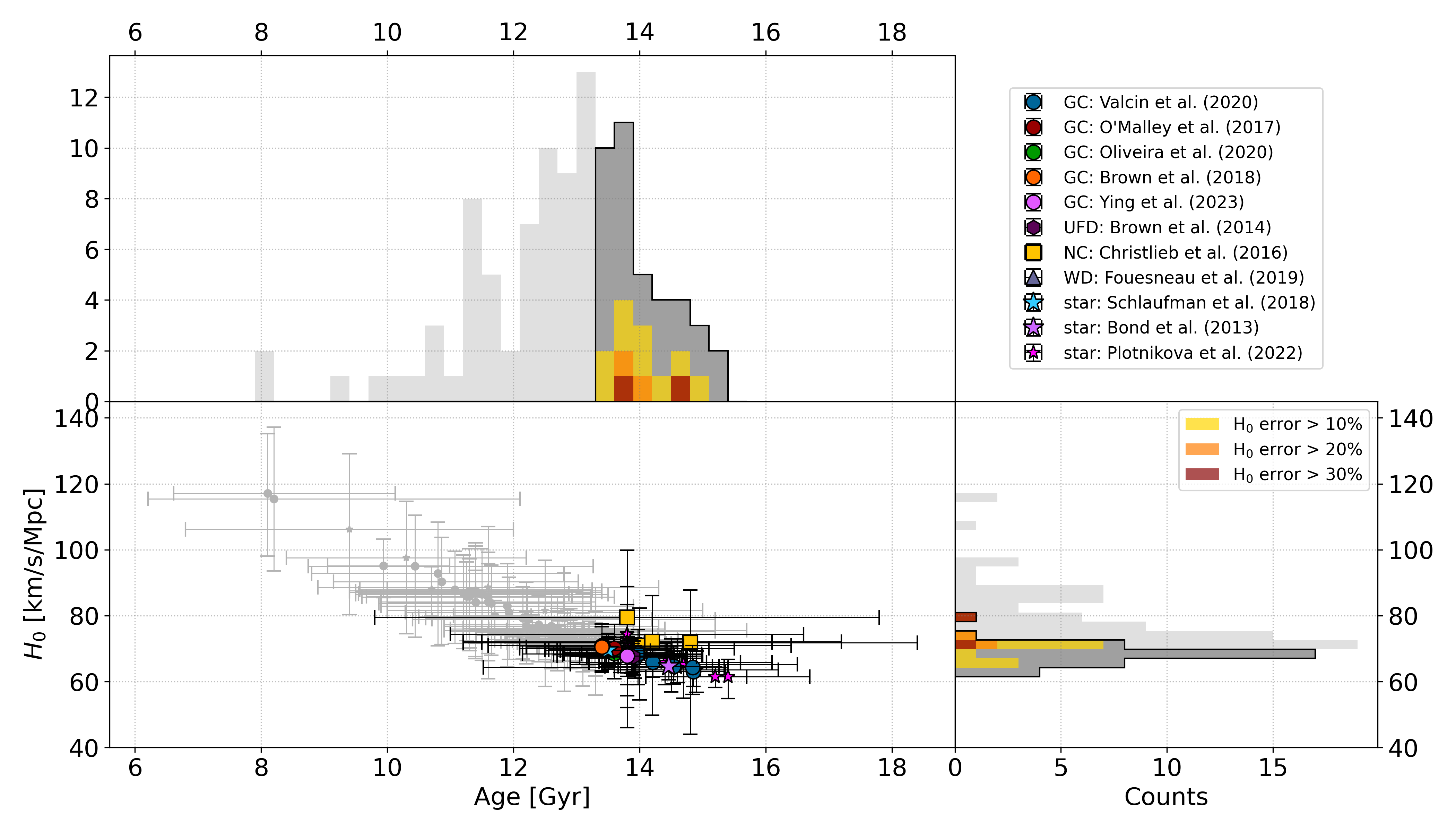}
\caption{Constraints on $H_0$ derived from the oldest stellar ages. In the bottom-left panel the distribution of the derived $H_0$ from the individual ages is shown, highlighting with colored points the 39 data used in our analysis (age $>13.3$ Gyr) on top of the data sample without any cut on age (grey points). The error bars include the statistical and systematic errors provided in each work. The top and right histograms show the distribution of ages and $H_0$, respectively, with the lighter grey representing the entire distribution and the darker grey our selected sample. We also further highlighted the data with larger errors on $H_0$, color-coded according to the percentage error (red considering a percentage error on $H_0>$30\%, orange for $>$20\%, yellow for $>$10\%).}
\label{fig:H0ages}
\end{figure*}

\subsection{Individual ages}
As a first step, we analyzed the individual age estimates of each object presented in Sect.~\ref{sec:age_meas}, considering an age threshold $>13.3$ Gyr to select the oldest objects. This value has been chosen to select at least one object for each sample, in order to preserve the variety of age dating results obtained with different methods and samples, therefore mitigating the possible biases. In the context of the Hubble tension, this is a conservative choice because an older age threshold would have provided lower $H_0$ values. We verified that our main results are robust and do not significantly depend on this assumption, as discussed in more detail in Sect~\ref{sec:oldthresh}. Based on this approach, 39 objects older than 13.3 Gyr were selected, and the Hubble constant $H_0$ was estimated for each of them. 
Fig.~\ref{fig:H0ages} shows that $H_0\lesssim72$ km/s/Mpc for the majority of our data, with values typically in the range $63<H_0\;{\rm[km/Mpc/s]<72}$. By inspecting the posteriors, the highest values of $H_0$ are due to the largest uncertainties on the ages and the consequent asymmetric PDF (Fig.~\ref{fig:H0}). The cases with $\sigma_{H_0}/H_0>$30\% have a mean age error $\sigma_{\rm age}=4$ Gyr, noticeably larger than the average of the entire sample ($\sigma_{\rm age}=1.4$ Gyr). Instead, for the cases with $\sigma_{H_0}/H_0<$30\%, $<$20\% and $<$10\%, the highest values of $H_0$ are 74.4, 71.9 and 70.6 km/s/Mpc, respectively (see the histogram in Fig.~\ref{fig:H0ages}).

We also tested how $H_0$ can be constrained with the individual very oldest globular clusters with the smallest age errors. For NGC 6362 ($13.6 \pm 0.5$ Gyr) \citep{oliveira2020} and NGC 6779 ($14.9^{+0.5}_{-0.9}$ Gyr) \citep{valcin2020}, we obtain $68.5^{+2.9}_{-3.2}$ and $63.1^{+3.9}_{-4.5}$ km/s/Mpc, respectively. Taken at face value, this exercise highlights the importance of the oldest objects in the context of the Hubble tension. However, these two individual cases are clearly insufficient to place meaningful constraints. For this reason, we also follow another approach based on the average ages (see the next Subsection). 

\begin{figure}[t!]
    \centering
    \includegraphics[width=0.44\textwidth]{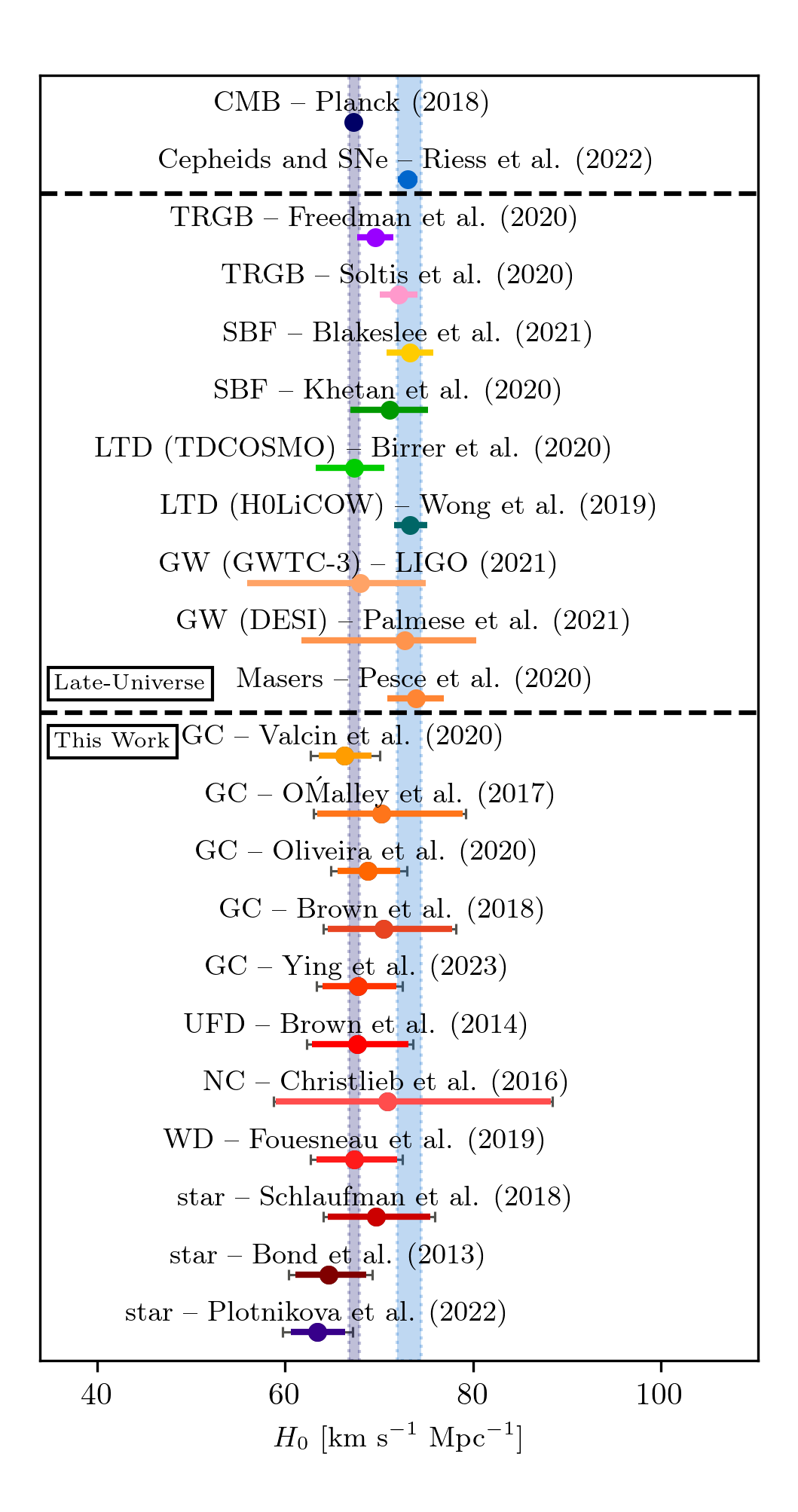}
    \caption{A comparison between the values of $H_0$ derived in this work and the ones available in the literature. \textit{Upper part}. The $H_0$ values from \cite{planck2020} and \cite{riess2022}. The vertical purple and blue shaded regions show the $\pm$1-$\sigma$ uncertainties of the two measurements in the entire plot. \textit{Central part}. The $H_0$ measurements obtained with different cosmological probes of the late-Universe: the tip of the red giant branch \citep[TRGB,][]{freedman2020,soltis2021}, surface brightness fluctuations \citep[SBF,][]{blakeslee2021,khetan2021}, lensing time delay \citep[LTD][]{birrer2020, wong2020}, gravitational waves \citep{LIGO2021,palmese2021}, and masers \citep{pesce2020}. \textit{Lower part}. Our estimates for each subsample in Sect.~\ref{sec:age_meas}. The inner thicker error bars show the uncertainty including the statistical and systematic errors for the age estimates. The outer thinner error bars show the total errors including also the additional systematic uncertainty derived from the adopted priors discussed in Sect.~\ref{sec:sim}.}
    \label{fig:H0all}
\end{figure}

\begin{table*}[t!]
\centering
\begin{tabular}{lccccc}
\hline
\hline
 & \# of & Mean Age & $H_0$  & $P(H_{0} \geq H_{0,{\rm Planck}})$ & $P(H_{0} \leq H_{0,{\rm SH0ES}})$ \\
Method & objects & [Gyr] & [km/s/Mpc]  & & \\
\hline
GC \citep{valcin2020} & 14 & 14.08$\pm$0.53 & $66.32^{+2.91}_{-2.75}$ & 0.35 & 0.99\\
GC \citep{omalley2017} & 2 & 13.49$\pm$1.4 & $70.28^{+8.64}_{-6.85}$ & 0.65 & 0.64\\
GC \citep{oliveira2020} & 2 & 13.57$\pm$0.85 & $68.82^{+3.43}_{-3.20}$& 0.67 & 0.90\\
GC \citep{brown2018} & 1 & 13.4$\pm$1.2 &  $70.52^{+7.25}_{-5.99}$ & 0.68 & 0.65\\
GC \citep{ying2023} & 1 & 13.80$\pm$0.75 & $67.76^{+4.10}_{-3.75}$ & 0.54 & 0.91\\
UFD \citep{brown2014} & 1 & 13.9$\pm$1 & 67.69$^{+5.46}_{-4.83}$ & 0.52 & 0.85\\
NC \citep{christlieb2016} & 4 & 14.17$\pm$2.5 & $70.89^{+17.37}_{-11.86}$ & 0.59 & 0.56\\
WD \citep{fouesneau2019} & 1 & 13.89$\pm$0.84 & $67.37^{+4.56}_{-4.02}$ & 0.50 & 0.91\\
Individual star \citep{schlaufman2018} & 1 & 13.5$\pm$1 & $69.66^{+5.81}_{-5.08}$ & 0.66 & 0.73\\
Individual star \citep{bond2013} & 1 & 14.46$\pm$0.8 & $64.67^{+3.99}_{-3.54}$ & 0.23 & 0.99\\
Very Metal Poor Stars \citep{plotnikova2022} & 11 & 14.73$\pm$0.59 & $63.45^{+2.93}_{-2.80}$ & 0.08 & 0.999\\
\hline
\end{tabular}
\caption{Constraints on the Hubble constant $H_0$ based on the average ages of the objects older than 13.3 Gyr present in each of the 10 independent samples. For each sample, we report the number of objects available, their mean age including statistical and systematic errors, and the estimated $H_0$. The last two columns report the probability that the estimated $H_0$ is respectively higher than the one obtained from Planck2020 \citep{planck2020} and lower than the value from SH0ES \citep{riess2022}.}
\label{tab:H0_data}
\end{table*}

\subsection{Average ages}
\label{sec:average_ages}
In order to minimize the potential bias induced by the larger age errors and to obtain more stringent constraints on $H_0$, we refined our analysis by averaging the age estimates (always keeping the oldest objects with ages $>13.3$ Gyr and separating the statistical and systematic contribution to the total error). Since each sample is characterized by its own systematic uncertainties, we decided not to average all data into a single age estimate. Therefore, for each of the 10 different samples reported in Sect.~\ref{sec:age_meas}, we estimated a mean age with an inverse-variance weighted average, adding a posteriori in quadrature the systematic error of each method as discussed in the corresponding paper, not to artificially reduce it with the averaging procedure. We analyzed these data with the same procedure described in Sect.~\ref{sec:method}, and the results are reported in Tab.~\ref{tab:H0_data}. 
We underline here that, wherever possible, we adopted the systematic errors reported in the original papers. However, in the cases where the contribution to the error budget due to systematic uncertainties was not explicitly and quantitatively reported \citep[e.g.][]{omalley2017, christlieb2016}, we adopted a conservative approach considering the systematic error to be the dominant contribution in the total error budget and taking as a reference the smallest total error (where most of the error should be driven by the systematic contribution). For instance, in the case of \cite{omalley2017}, where the most accurate ages have total errors (statistical plus systematic) of $\pm$1.3 Gyr, we adopted a systematic error of $\pm$1 Gyr. For \cite{christlieb2016}, we assumed $\pm$2 Gyr as the systematic uncertainty.
We found that $63.5<H_0\;{\rm[km/Mpc/s]<70.9}$, with errors around 2.8 km/s/Mpc in the best case and around 14.1 km/s/Mpc in the worst one. If the systematic errors due to the choice of our priors (see Sect.~\ref{sec:sim}) are also added, the total uncertainties slightly increase to 3.7 and 14.3 km/s/Mpc, respectively. 

The results are presented in the framework of the Hubble tension showing, for each subsample, the average probability (weighted with the sample size) of each $H_0$ to be larger than the Planck value \citep{planck2020} or smaller than the SH0ES one \citep{riess2022}. 

The results indicate an average probability of 90.3\% of the Hubble constant to be $H_0<73.0$ km/s/Mpc (weighted on the number of data points in each sample), with a minimum value of 56\% and a maximum value of 99.9\%. Instead, the average probability to have $H_0>67.4$ km/s/Mpc is 35.7\%, with a minimum value of 8\% and a maximum value of 68\%. If also the conservative systematic error due to the choice of priors (Sect.~\ref{sec:sim}) is added, the average probabilities discussed above change only by a few percent. In Fig.~\ref{fig:H0all}, we compare our estimates with other $H_0$ constraints from the literature including a collection of $H_0$ measurements obtained with late-Universe probes.

All our results based on the oldest stellar ages, indicate a statistical preference for a value of $H_0$ smaller than the SH0ES constraint and more compatible with the Planck2020 results, even if the current error bars are still quite large and dominated by systematics. In Sect. ~\ref{sec:oldthresh} we explore how our conclusions are affected by the choice of the adopted age threshold.

\subsection{How old is too old?}
\label{sec:oldthresh}
The most stringent constraints on $H_0$ come from the oldest objects in the present-day Universe. However, the definition of “oldest” depends on the arbitrary choice of an age threshold that can be affected by statistical biases. For instance, choosing a small number of objects with the very oldest ages might bias the analysis toward fluctuations in the age distribution. 
In our work, we selected ages older than 13.3 Gyr because this allowed us to keep at least one object in each sample and to maximize the variety of the age estimate methods. In other words, a threshold of $>$13.3 Gyr was the best trade-off between selecting the “oldest” objects, maximizing the variety of object types and the different methods used to derive their stellar ages, and therefore minimizing potential biases driven by specific classes of objects.

Here, we discuss how other choices may influence the final results on $H_0$. In particular, we asked ourselves the question: {\it how old is ``too old''?}. We followed three approaches to address this question:

{\it (i)} we verified whether our chosen threshold was so high that it selected statistical fluctuations in our dataset,

{\it (ii)} we checked how our result changed by lowering the adopted age threshold, and

{\it (iii)} we tested alternative criteria to select the oldest object in the Universe.

\begin{figure*}
\centering
\includegraphics[width=0.45\textwidth]{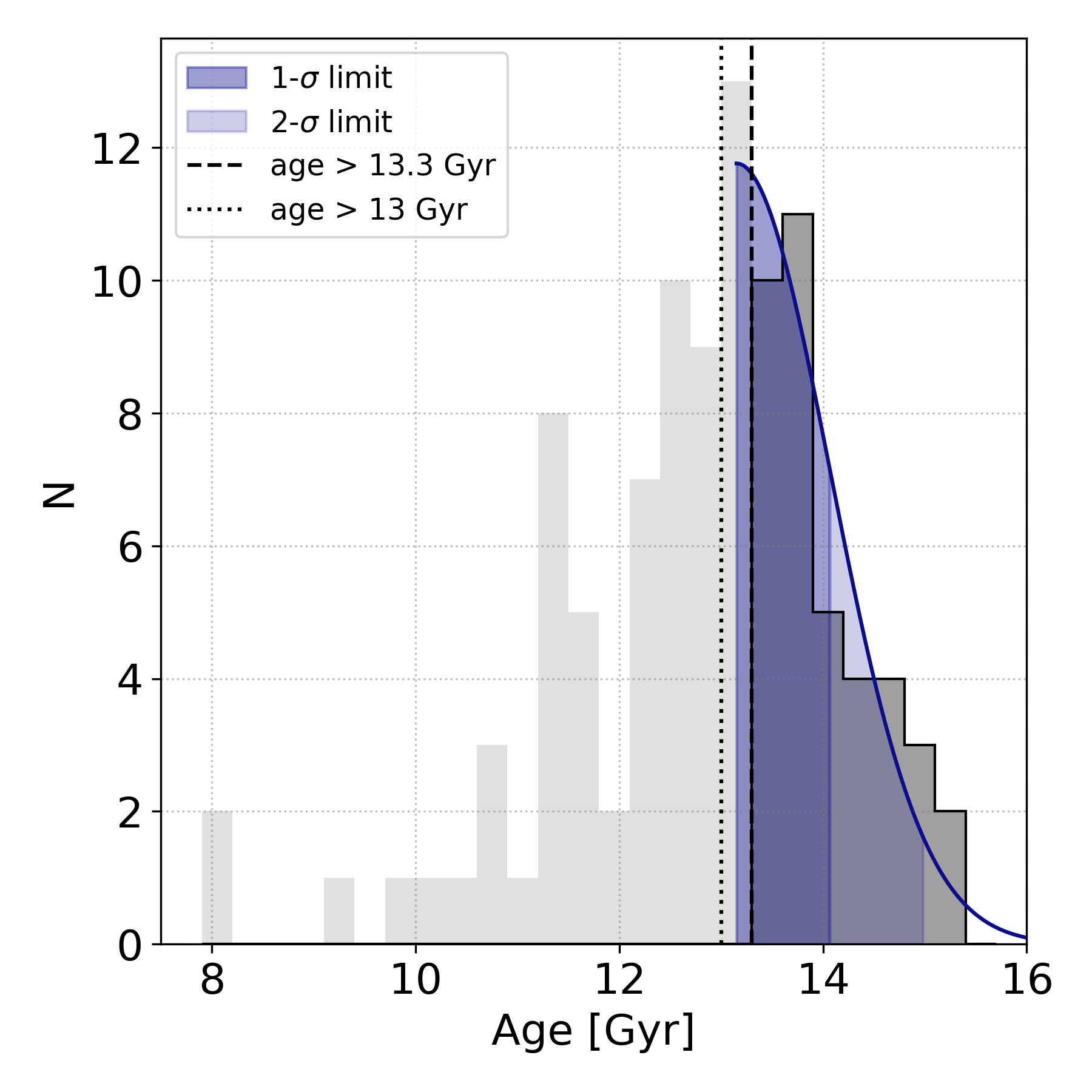}
\includegraphics[width=0.45\textwidth]{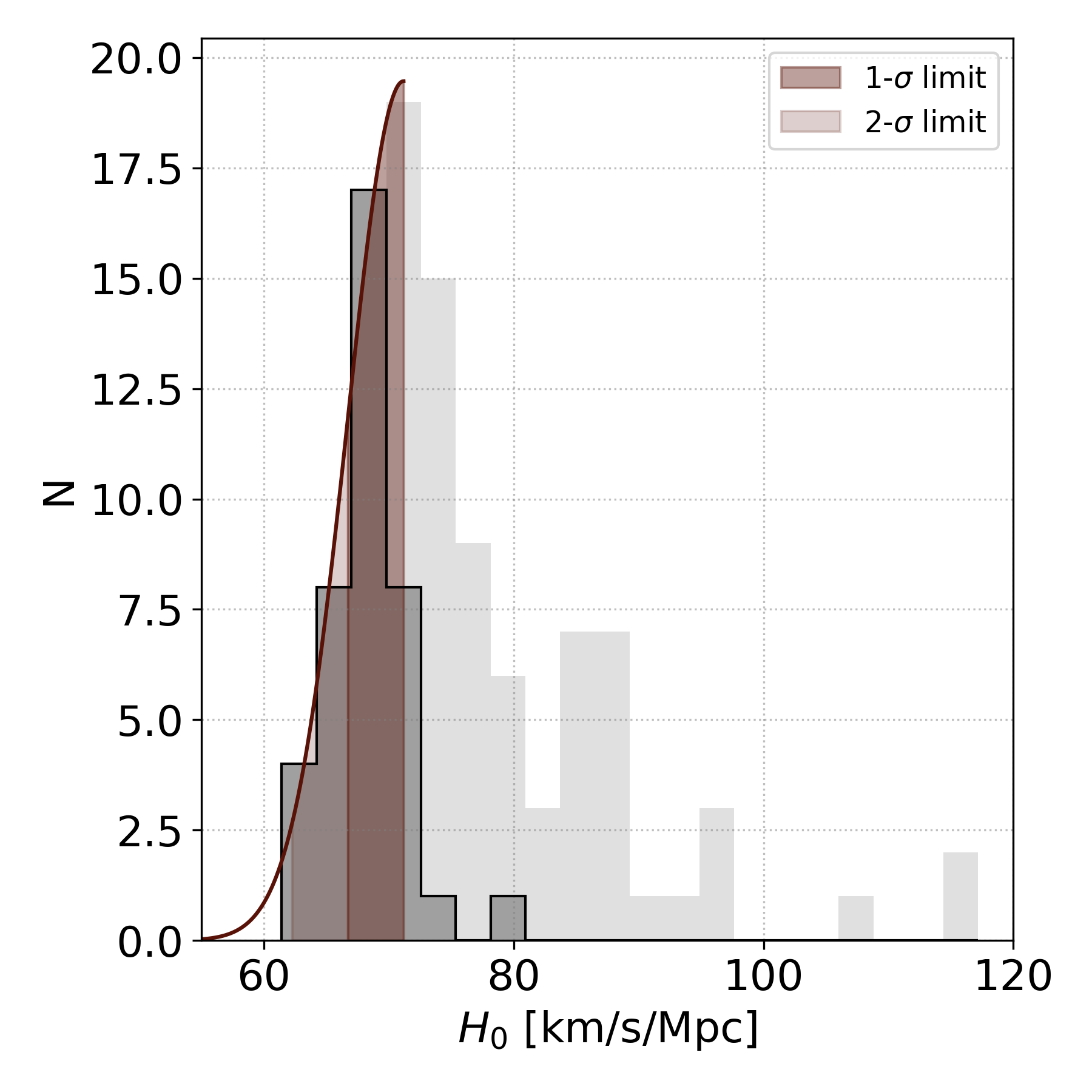}
\caption{\textit{Upper panel.} The distribution of ages in our sample. The lighter and darker gray histograms show the overall sample and the objects with age$>13.3$ Gyr, respectively. The vertical lines show the cuts adopted in the main text (age$>13.3$ Gyr) and the one investigated in this section (age$>13$ Gyr). The blue line is the one-sided Gaussian fit to the edge of the age distribution, highlighting in dark blue and light blue the 1-$\sigma$ and 1-$\sigma$ ranges, respectively. This demonstrates that the $1\sigma$ of the upper edge of the age distribution is larger than both the adopted cuts. \textit{Lower panel.} The distribution of estimated values of $H_0$. In this case, the lower envelope of the $H_0$ distribution is fitted with a one-sided Gaussian, corresponding to the older ages, showing in dark and light red the 1-$\sigma$ and 1-$\sigma$ ranges, respectively. In this case, we find that the $1\sigma$ of the lower edge of the $H_0$ distribution obtained from the individual age is compatible with the results we obtained from the average analysis in Sect.~\ref{sec:average_ages}.}
\label{fig:H0ages_gaussfit}
\end{figure*}

{\it (i)} As a first test, we studied the global age distribution of all the objects presented in the various papers to characterize the upper envelope of the distribution\footnote{Here, we just consider a lower cut in age $>$ 8 Gyr, since it would be not useful to determine the upper edge of the distribution.}. These data are presented in the left panel of Fig.~\ref{fig:H0ages_gaussfit}. We found that the age distribution can be modeled as an asymmetric distribution with a distinctive peak around age$\sim$13 Gyr. We modeled the upper edge of the distribution with a one-sided Gaussian, which is also shown in the figure, and calculated its width $\sigma$. We find that the threshold chosen in our analysis (age$>$13.3 Gyr) corresponds to $<1\sigma$ of the distribution of the upper envelope of our sample (shown as the darker shaded area in the left panel of Fig.~\ref{fig:H0ages_gaussfit}), showing in this way that the threshold $>$13.3 Gyr does not select high statistical fluctuations in our data. Note that also the distribution of the derived Hubble constant provides a similar piece of information. If we study the lower edge of the $H_0$ distribution (corresponding to the older ages; see the right panel of Fig.~\ref{fig:H0ages_gaussfit}), we find that it can also be modeled by a one-sided Gaussian, where the $1-\sigma$ limit is $H_0=66.7$ km/s/Mpc, which is in very good agreement with the mean values estimated from the various samples derived in Sect.~\ref{sec:average_ages}.

{\it (ii)} Next, we also investigated how much the main results of our analysis change by lowering the threshold from 13.3 Gyr to 13 Gyr. This clearly goes against the purpose of this work to exploit the oldest objects as cosmological probes, but it allowed us to test the dependence of our findings on the chosen threshold. We decided not to choose a younger age threshold since we verified in Fig.~\ref{fig:H0ages_gaussfit} that this is the peak of the age distribution, and considering younger ages would have biased the results in the opposite direction by including objects that are not any more representative of the upper edge of the distribution. We then repeated the analysis of Sect.~\ref{sec:average_ages} with a cut of $>$13 Gyr, and the results are reported in Tab.~\ref{tab:H0_data_lowerlim}. As expected, we find that the mean ages decrease and, consequently, the derived values of $H_0$ move towards higher values. However, the change is limited. In particular, we find that the derived Hubble constant is more likely to be in a range between the SH0ES and Planck values, with still a slightly larger preference to be smaller than the SH0ES value. In fact, the weighted probability of $H_0$ being smaller than the SH0ES values changes from 90.3\% to 89.9\%, whereas the probability to be higher than the Planck value changes from 35.2\% to 42.7\%.

{\it (iii)} Finally, we also explored a different approach to select the oldest objects as those with the lowest metallicity based on the well-known anticorrelation between age and metal abundance. However, we note that this method would have been applicable homogeneously only to a subsample of our dataset. For this reason, we tested this approach using directly the data of \cite{valcin2020} who defined their oldest GC sample by selecting the objects with metallicity [Fe/H]$< -1.5$. With such a criterion, they obtained an age$=13.32\pm0.1(stat.)\pm0.5(syst.)$ Gyr. If we use their age estimate to derive the Hubble constant, we obtain $H_0= 70.07^{+3.18}_{-2.94}$ km/s/Mpc. In \cite{valcin2021}, they also revised and updated the systematic error estimate to $\sigma_{stat}(H_0)=0.23$ km/s/Mpc, and if we use this estimate we obtain $H_0=69.87^{+1.94}_{-1.83}$ km/s/Mpc. Both values are slightly larger than the estimate we provided in our work, but still compatible within the $1-\sigma$ errors, and do not change our general findings.

\begin{table*}[t!]
\centering
\begin{tabular}{lccccc}
\hline
\hline
 & \# of & Mean Age & $H_0$  & $P(H_{0} \geq H_{0,{\rm Planck}})$ & $P(H_{0} \leq H_{0,{\rm SH0ES}})$ \\
Method & objects & [Gyr] & [km/s/Mpc]  & & \\
\hline
GC \citep{valcin2020} & 19 & 13.87$\pm$0.53 & $67.30^{+2.99}_{-2.79}$ & 0.47 & 0.98\\
GC \citep{omalley2017} & 5 & 13.27$\pm$1.1 & $71.24^{+7.26}_{-6.12}$ & 0.71 & 0.62\\
GC \citep{oliveira2020} & 2 & 13.57$\pm$0.85 & $68.82^{+3.43}_{-3.20}$& 0.67 & 0.90\\
GC \citep{brown2018} & 1 & 13.4$\pm$1.2 &  $70.52^{+7.25}_{-5.99}$ & 0.68 & 0.65\\
GC \citep{ying2023} & 1 & 13.80$\pm$0.75 & $67.76^{+4.10}_{-3.75}$ & 0.54 & 0.91\\
UFD \citep{brown2014} & 1 & 13.9$\pm$1 & 67.69$^{+5.46}_{-4.83}$ & 0.52 & 0.85\\
NC \citep{christlieb2016} & 4 & 14.17$\pm$2.5 & $70.89^{+17.37}_{-11.86}$ & 0.59 & 0.56\\
WD \citep{fouesneau2019} & 1 & 13.89$\pm$0.84 & $67.37^{+4.56}_{-4.02}$ & 0.50 & 0.91\\
Individual star \citep{schlaufman2018} & 1 & 13.5$\pm$1 & $69.66^{+5.81}_{-5.08}$ & 0.66 & 0.73\\
Individual star \citep{bond2013} & 1 & 14.46$\pm$0.8 & $64.67^{+3.99}_{-3.54}$ & 0.23 & 0.99\\
Very Metal Poor Stars \citep{plotnikova2022} & 16 & 14.40$\pm$0.57 & $64.80^{+2.97}_{-2.77}$ & 0.18 & 0.998\\
\hline
\end{tabular}
\caption{Constraints on the Hubble constant $H_0$ based on the average ages of the objects older than 13 Gyr present in each of the 10 independent samples. For each sample, we report the number of objects available, their mean age including statistical and systematic errors, and the estimated $H_0$. The last two columns report the probability that the estimated $H_0$ is respectively higher than the one obtained from Planck2020 \citep{planck2020} and lower than the value from SH0ES \citep{riess2022}.}
\label{tab:H0_data_lowerlim}
\end{table*}

\section{Accuracy matrix and prospects}
\label{sec:forecasts}
\begin{figure}[t!]
    \centering
    \includegraphics[width=0.5\textwidth]{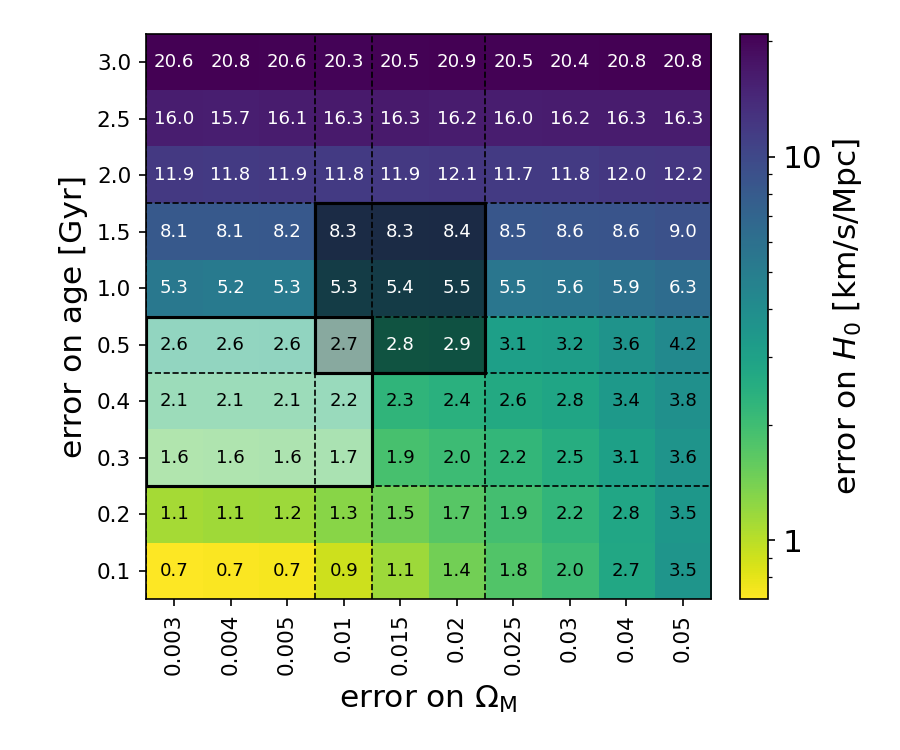}
    \caption{The expected errors on $H_0$ for the {\it reference case} (see Sect.~\ref{sec:cosmoconstraints}; age=13.5 Gyr and $11<z_f<30$) as a function of the uncertainty on $\Omega_{\rm M}$ (x-axis) and on the age of the oldest stellar objects (y-axis). Given the pair $(err(\Omega_{\rm M},err({\rm age}))$, the derived error on $H_0$ (in km/s/Mpc) is shown in each square. The darker shaded square indicates the range of errors currently spanned in this paper. The lighter shaded square shows the improvement expected from the higher accuracies that could be reasonably obtained in the near future.}
    \label{fig:err_matrix}
\end{figure}

The results presented in the previous section show the high potential of the oldest stars as cosmological probes. The constraints on $H_0$ can become more stringent with higher accuracy of stellar ages and $\Omega_{\rm M}$. We used the workflow presented in previous sections to construct a matrix that shows how the accuracy of $H_0$ depends on the errors on stellar ages and $\Omega_{\rm M}$. The uncertainty on the age in Fig.~\ref{fig:err_matrix} is the total one, i.e. including statistical and systematic errors.
First, we notice that the uncertainty on the age dominates the error budget of $H_0$, whereas the uncertainty on $\Omega_{\rm M}$ has a subdominant effect. The minimum uncertainty $\sigma_{H_0} \sim$2.5 km/s/Mpc currently attainable (darker square) is larger by a factor of 2-4 than the most accurate estimates of $H_0$ available to date \citep{planck2020,riess2022}. However, the matrix also shows that significant improvements are expected in case the error on $\Omega_{\rm M}$ is reduced to $\sim$0.003 \citep[e.g. with the Euclid mission, see][]{amendola2018} and the total error on the age decreases to $\sim$0.3 Gyr. 

More accurate ages could be achieved by increasing the sample size (i.e. minimizing the statistical error) and by further reducing the systematics (e.g. \citep[e.g.][]{valcin2021,wu2022}. This would allow us to reach an accuracy on $H_0$ of the order of $\lesssim$1.5 km/s/Mpc that could play a decisive role in the Hubble tension.
\section{Summary and outlook}
The oldest stars in the present-day Universe play a key role as independent cosmological probes. In this work, we collected a sample of stellar objects for which state-of-the-art age estimates were available in the literature to revisit their potential to constrain the Hubble constant. The sample includes different types of objects (globular clusters, individual stars, white dwarfs, ultra-faint and dwarf spheroidal galaxies) whose ages were estimated with independent methods taking into account statistical and systematic uncertainties. The main results of this work can be summarized as follows.
\begin{itemize}
\item We built a Bayesian framework to constrain the Hubble constant exploiting the age of the oldest stars. We adopted a flat $\Lambda$CDM model, assuming a flat prior on the formation redshifts ($11<z_{\rm f}<30$) and a Gaussian prior on $\Omega_{\rm M}=0.30 \pm 0.02$ based on late-Universe probes independent of the CMB constraints. This prior choice has been estimated to affect our error estimate at most with a systematic error of $\sigma_{syst,\;prior}(H_0)=2.37$ km/s/Mpc, which is, however, highly conservative because the observational constraints significantly limit the actual range of priors.
\item We selected 39 objects with ages older than 13.3 Gyr and, for each object, we estimated the Hubble constant. The distribution of $H_0$ is concentrated in the range of $63<H_0\;{\rm[km/Mpc/s]<72}$, with a preference for low values of $H_0$ if the most accurate estimates are selected. Although the current age uncertainties of individual objects do not allow stringent constraints on $H_0$, the results clearly show the key role of the oldest objects as independent cosmological probes.
\item If the ages are averaged and analyzed independently for each subsample, we derived more stringent constraints that imply a high probability (90.3\% on average) of $H_0$ to be lower than the SH0ES value, and indicate that the ages of the oldest stars are more compatible with the Planck2020 estimate. 
\item We constructed an ``accuracy matrix'' to assess how the constraints on $H_0$ can be tightened by increasing the accuracy of stellar ages and $\Omega_{\rm M}$. Should the systematic errors on stellar ages be reduced to $\lesssim 0.3-0.4$ Gyr, the accuracy of $H_0$ would increase to $\sim$1-2 km/s/Mpc and become fully competitive with the other cosmological probes shown in Fig.~\ref{fig:H0all}.
\end{itemize}
The results presented in this work show the high potential and a bright future for the oldest stars as cosmological probes. Several improvements can be expected thanks to massive spectroscopic surveys of extremely/very metal-poor stars (e.g. PRISTINE, WEAVE) combined with the parallax information provided by \textit{Gaia}.
In this regard, the recent results on the metal-poor GC M92 \citep{ying2023} are also indicative of the improvements expected in the estimate of the absolute ages of GCs with accuracies high enough to allow for cosmological applications with a full control of the systematic uncertainties. It is indeed remarkable that in the case of M92 the error budget is dominated by the distance of this GC and not by the stellar evolution models.
Moreover, spectroscopy with extremely large telescopes will allow us to apply nucleochronometry to larger samples and possibly reduce the age uncertainties to $\sim$0.3 Gyr \citep{wu2022}. Another promising opportunity will be offered by further studies of dwarf and ultra-faint galaxies in the Local Group. 
The imminent deep and homogeneous data obtained with space-based imaging (e.g JWST, \citealt{weisz2023a}; \textit{Euclid}, \citealt{laureijs2011}) will allow us to reconstruct the star formation histories of these galaxies with higher fidelity and therefore to derive the ages of the oldest stellar populations (see for instance \citealt{weisz2023b} for a recent example of this promising approach). In parallel, improved stellar evolution and white dwarf cooling models will likely reduce the systematic uncertainties on age dating.

These advances will enhance the constraining power of the oldest stars in cosmology and their full exploitation in synergy with the forthcoming results expected from \textit{Euclid} and other cosmological surveys. Moreover, the role of the oldest stars will not be limited only to $H_0$, but also crucial for other cases in cosmology and fundamental physics. For instance, significant constraints could be placed to test the early dark energy (EDE) scenarios, invoked to mitigate the Hubble tension by adjusting the sound horizon. In that case, the younger age of the universe required in EDE models (e.g. \citealt{smith2022, poulin2023}) seems to be incompatible with the ages of the oldest objects in the present-day Universe, proving how this data could be the optimal testbed to discriminate between different cosmological models.

\begin{acknowledgments}
M.M. and A.C. acknowledge the grants ASI n.I/023/12/0 and ASI n.2018-23-HH.0. A.C. acknowledges the support from grant PRIN MIUR 2017 - 20173ML3WW\_001. M.M. acknowledges support from MIUR, PRIN 2017 (grant 20179ZF5KS). The authors are grateful to the referee, Adam Riess, and Raul Jimenez for the constructive discussion and comments.
\end{acknowledgments}

\software{\textsc{emcee} \citep{foreman2013},
          \textsc{ChainConsumer} \citep{hinton2016},  
          \textsc{Matplotlib} \citep{hunter2007},
          \textsc{Numpy} \citep{harris2020},
          \textsc{plot1d} \url{https://github.com/Pablo-Lemos/plot1d}.
          }


\bibliography{bib}{}
\bibliographystyle{aasjournal}

\end{document}